\documentclass{kluwer}    

\newdisplay{guess}{Conjecture}

\begin{document}                                                                                   
\begin{article}
\begin{opening}         
\title{The moon and the origin of life\footnote{Earth, Moon and Planets
85/86, 61 (2001)}} 
\author{C.R.  \surname{Benn}}  
\runningauthor{C.R. Benn}
\runningtitle{The moon and the origin of life}
\institute{Isaac Newton Group, Apartado 321, 38700 Santa Cruz de La Palma,
Spain}

\begin{abstract}
Earth is unusual in bearing life, and in having a large moon.
A number of authors have suggested a possible connection between the two,
e.g. through lunar stabilisation of the earth's obliquity,
or through the effects of the oceanic tides.
The various suggestions are reviewed.

\end{abstract}

\end{opening}

\section{Introduction}
The properties of the universe that we observe must be consistent
with the evolution of carbon-based life within it.
This observational selection effect is known as the 
{\it weak anthropic principle} (Barrow \& Tipler 1986).
It has been invoked to explain 
a number of otherwise unlikely coincidences such as the
nuclear resonance that allows carbon to form
in stellar interiors, 
the big numbers coincidence (ratio of
strengths of electromagnetic and gravitational forces $\sim$ 10$^{40}$
$\sim$ current size of the observable universe in proton diameters),
and, more recently,
the smallness of the cosmological constant (Efstathiou 1995)
and the amplitude of the primordial density fluctuations which seeded
the growth of galaxies and clusters of galaxies
(Tegmark \& Rees 1998).

A similar selection effect will apply in our local
astrophysical environment.
Clearly,
the luminosity and lifetime of the sun, and the shape and size of
the earth's orbit, 
must be such as to maintain the earth's surface,
for a long period, at a temperature
suitable for the evolution of organic life
(e.g. Kasting et al 1993).
However, our solar system may also be atypical in other respects:
\newline $\bullet$
The sun's metallicity is unusually high for its age
(Whittet 1997, Gonzalez 1999), perhaps reflecting
the higher probability of planetary systems being associated
with high-metallicity parent stars.
\newline $\bullet$
The sun's luminosity may be unusually stable (Gonzalez 1999).
\newline $\bullet$
Jupiter's role in ejecting comets from the solar system
could have been crucial in protecting the young earth from life-inhibiting
impacts (Wetherill 1995).
Solar systems with Jupiter-like planets at similar radii may thus not be 
typical.
\newline $\bullet$
The sun may be orbiting the galaxy 
close to the co-rotation circle (Mishurov \& Zenina 1999),
which minimises the number of spiral-arm crossings, and 
consequent disruption of the solar system (e.g. by nearby supernovae,
tidal effects).

Although it's possible that
none of the above features of our solar system is {\it essential} to the 
evolution of life
on earth, the probability of our observing them is enhanced
if they increase the probability that intelligent life will develop, i.e.
it would not be surprising to observe any feature $F$ if its
a priori probability, $p_F$, satisfies:

\hspace*{15mm} $p_{F}>p_{life}(0)/p_{life}(F)$ \hfill (1)

where $p_{life}(0)$ and $p_{life}(F)$ are the probabilities of intelligent
life evolving respectively in the absence and 
in the presence of feature $F$.

Several authors (e.g. Butler 1980, Comins 1993) 
have suggested that there might be a link between our large moon,
arguably an a priori unlikely feature of the earth's environment,
and the evolution of life on earth, i.e. that:

\hspace*{15mm} $p_{moon}>p_{life}(0)/p_{life}(moon)$ \hfill (2)

All 3 parameters in this inequality are unknown.
Below we consider what is known about the origin of the moon, and the
origin of life, and how the former might affect the latter.

\section{The origin of the moon}
The earth's moon is unusual amongst solar-system satellites in having
relatively 
high mass and in dominating the total angular momentum 
of the earth-moon system.
It is also unusually `dry',
lacking volatile elements such as K and Bi.
These characteristics, particularly the high angular momentum,
posed serious problems for the three main hypotheses for the
origin of the moon until the mid 1980s: 
capture, co-formation and fission (Hartmann et al 1985).
There is now a consensus (Taylor 1992) that none of these three hypotheses
is tenable, and that the earth-moon system probably formed
when a planetesimal $\sim$ 0.1 - 0.2 
earth masses underwent a grazing collision with the proto-earth,
soon after the formation of the solar system 4600 Myr ago.
In this widely-accepted 
`Big Splash' scenario, the metallic core of the impacting
body sank to join the earth's, and some of the shattered mantle
reassembled in orbit to form the moon.
Estimates of the probability ($p_{moon}$)
of the earth acquiring such a large moon have been boosted by this new perspective, but e.g. Ringwood (1990) and Lissauer (1997) argue 
that only a narrow range of initial conditions could have resulted 
in a large moon in earth orbit.

\section{The origin of life}
Life on earth probably dates back at least 3500 Myr,
i.e. to within a few 100 Myr of the formation of the earth's crust
4200 - 4000 Myr ago.
The oceans at that time would have been a weak solution of organic
precursors (amino acids, pyrimidines etc.), 
formed in situ, or in the comets that
may have provided much of the water (Chyba 1990, 
Lazcano-Araujo \& Or\'{o} 1981).
How a self-replicating system evolved from this primordial soup
has been the subject of much speculation and
laboratory work, reviewed by Or\'{o} et al (1990) and
Deamer \& Fleischaker (1994).
One common theme to emerge is the importance of concentrating
the soup to encourage polymerisation, e.g. in tidal pools which repeatedly
dry out under the sun,
echoing Darwin's (1871) speculation about the origin of life in
``some warm little pond with all sorts of ammonia and phosphoric salts''.
For example, Or\'{o} et al (1990)
note that the best contemporary laboratory models
of pre-cellular systems are liposomes
(phospholipid vesicles), and
encapsulation of DNA within liposomes 
has been achieved by dehydration-hydration cycles similar to those
occurring in intertidal pools (Deamer \& Barchfeld 1982).

Once self-replication is established, evolution to more complex systems
can proceed through Darwinian selection.
{\it Intelligent} life is not an inevitable end product, and
the fact of our existence places no 
constraint on the probability
($p_{life}(0)$, or $p_{life}(moon)$)
of intelligent life evolving on an earthlike planet.
It might be close to 1, but it might just as well be
10$^{-30}$
(implying no other life within a Hubble radius).
The long intervals between critical events in the evolution of life on
earth suggest that evolution to intelligent life
is unlikely to happen much faster on
other earthlike planets.
The key requirements for the evolution of organic life on an
earthlike planet are thus a source of organic precursors,
solid surfaces where the precursors can condense,
long-term maintenance of temperature within a range suitable
for organic reactions,
and protection from hazards (e.g. impacts).

\section{Possible influence of the moon on the evolution of life}
Given on the one hand the wide-ranging consequences 
(compositional, gravitational) of the
earth having a large moon, and on the other
the stringent requirements for the origin and  
evolution of life, it would perhaps be surprising if 
the former had {\it not} significantly affected the latter.
A number of specific suggestions have been made:

{\bf (1) Stabilisation of the earth's obliquity}
\\
Small changes in the earth's orbit and orientation
probably drive climatic change
(Milankovitch theory, Imbrie 1982), 
and small changes of the earth's obliquity
(angle of spin axis with the perpendicular to the orbital plane)
of $\sim$ 1$^o$ could have triggered recent ice ages.
Several authors (e.g. Goldsmith \& Owen 1980, Verschuur 1989)
have noted that a large moon would benefit life by stabilising the
earth's obliquity, and thus climate, and
Laskar, Joutel \& Robutel (1993) and de Surgy \& Laskar (1997)
confirmed that
in the absence of the moon, large and chaotic variations of obliquity 
would have occurred.
Mercury and Venus have been stabilised by 
tidal dissipation (they spin very slowly),
but Mars, which has no large moon, undergoes chaotic variations
of obliquity in the range 0 - 60$^o$
(Laskar \& Robutel 1993).
Stabilisation of the earth's obliquity might not have been crucial for
the origin of life, but could have been for the evolution of life
on land.

{\bf (2) Elimination of the primordial atmosphere}
\\
Cameron \& Benz (1991) and Taylor (1992) pointed 
out that the giant impact which created the
earth-moon system would have stripped the earth of its thick primordial
atmosphere, which might otherwise have developed as has that of Venus,
rendering the surface of the planet too hot for organic life.
On the other hand, the earth's atmosphere may be thinner simply
because most of the CO$_2$ is locked up in carbonate rocks.

{\bf (3) Generation of the earth's magnetic field}
\\
The earth's magnetic field partially shields the molecules 
of life from the destructive effects of cosmic rays (although this
may not be important for submarine life).
Compared to other solar-system bodies, the
earth's magnetic field is unusually strong for its angular momentum
(though the mechanisms are probably different for different 
bodies).
Pearson (1988) suggested that this anomaly might be due to
the prolonged heating of the earth's core 
following the impact that created the earth-moon system.

{\bf (4) Generation of large tides}
\\
As noted above, a common theme in speculations about the origin of the first
self-replicating system
is the importance of concentrating the weak solution of organic
molecules in the primordial sea, to encourage polymerisation.
The possible role of tidal pools, which repeatedly dry out under the sun,
has been stressed by many authors. 
Although the amplitude of the tides raised by the moon is currently not
much larger than that of the
tides raised by the sun, the moon was probably
much closer to the earth at the time of the origin of life (Chyba 1990),
and the tides raised would have been correspondingly larger,
allowing tidal pools with a much larger total
area to be subjected to wetting/drying cycles
(e.g. Verschuur 1989, Gribbin \& Rees 1990).

{\bf (5) Generation of longer-period tides}
\\
The length of time allowed for intertidal pools to dry out under
the sun before being refilled may also have been important.
With longer wetting/drying cycles,
the probability of long sequences of chemical reactions
taking place is increased
(the energies involved in organic reactions are small,
and they proceed slowly).
It's possible that at the time of the origin of life, 
the moon was not actually much closer to earth
than it is at present (Williams 1989, Taylor 1992).
One may then speculate (Rood \& Trefil 1981)
that beating between lunar and solar tides
to give neap/high tides at a longer interval (as observed at present)
was important, although of course
longer intervals
could also be achieved by other means e.g. through seasonal
effects.
The hypothesis has an interesting corollary.
The condition for such beating is that the strengths of the
tides raised by the
sun and by the moon, which are 
$\propto$ density $\ast$ (angular diameter)$^3$, are similar.
The mean densities of planetary bodies and main-sequence stars both
happen to be $\sim$ atomic (Carr \& Rees 1979),
so angular diameter $\propto$ (strength of tide)$^{1/3}$ i.e.
the condition for long-period tides happens to 
imply similar angular diameters of the sun and moon, as 
observed.\footnote{Curiously, this has a literary antecedent;
in Martin Amis' novel
{\it London Fields} (Amis 1989) appears
the line `Perhaps that was the necessary
condition of planetary life: your sun must fit your moon'.}

With our current level of understanding of the origin and evolution
of life, the above five hypotheses remain speculative.
However, the variety of suggested mechanisms (and more than one could
be important) attests to
the far-reaching consequences of the earth having a large moon.
Some of these consequences
inevitably impinge on events critical to the development
of life on earth.
Thus it cannot be assumed 
that this unusual feature of our environment
is {\it not} anthropically selected.

\section{Conclusions}
In studies of the solar system, as in cosmology, we are dealing
with a unique example (so far), and we must beware the effects on our
observations of anthropic selection.
The possibility that the presence of the moon has affected the
origin or evolution of life through one of the mechanisms noted above
implies that:
\newline
(1) hypotheses about the origin of the moon cannot be judged
solely on the basis of a priori likelihood (they need only
satisfy equation 2);
\newline
(2) large moons may be useful pointers in the search for 
life-bearing planets.

More generally,
caution must be exercised in interpreting, and generalising from,
unusual features of our local astrophysical environment;
they may turn out to be anthropically selected, and atypical.
Indeed, the earth and its environment may be very special
(Ward \& Brownlee 2000), and this might explain the puzzling
lack of evidence for intelligent life elsewhere in the universe
(Tipler 1980, Wesson 1990).

\theendnotes

\end{article}
\end{document}